\documentclass[aps,nofootinbib,showpacs,twocolumn%,floatfix
]{revtex4}
\usepackage{axodraw}
\usepackage{graphicx}
\usepackage{epsf,amsfonts,amssymb,amsbsy}
\usepackage[mathscr]{eucal}

\begin{document}
\title{Isospin breaking in scalar and pseudoscalar channels of
radiative $J/\psi$-decays}
\author{V.V.Kiselev}
\email{Valery.Kiselev@ihep.ru}
%\fax{}
\affiliation{Russian State Research Center ``Institute for High
Energy Physics'', %\\
Pobeda 1, Protvino, Moscow Region, 142281, Russia\\ Fax:
+7-4967-744937}
\begin{abstract}
    In the framework of simple assumption on factorizing a
    mixing of vector state with isoscalar components in effective amplitudes
    of isospin breaking caused by the electromagnetic quark
    current, a branching fraction of radiative $J/\psi\to a_0$ transition is
    evaluated at the level of $3\cdot 10^{-3}$.
    % due to the isospin breaking by the electromagnetic quark
    %current.
\end{abstract}

\maketitle

%%%%%%%%%%%%%%%%%%%%%%%%%%%%%%%%%%%%%%%%%%%%%%%%%%%%%%%%%%%%%%%%%%
%\documentclass{revtex4}%[11pt]{article}
%\usepackage{axodraw}
%\textwidth=158mm \textheight=230mm \voffset=-1.5cm
%\begin{document}
%\title{\bf\large Isospin breaking in scalar and pseudoscalar channels of
%radiative $J/\psi$-decays}
%\author{V.V.Kiselev}
%\date{}
%
%\maketitle

\section{Introduction} 
\begin{figure*}[th]
\begin{center}
\includegraphics[width=14cm]{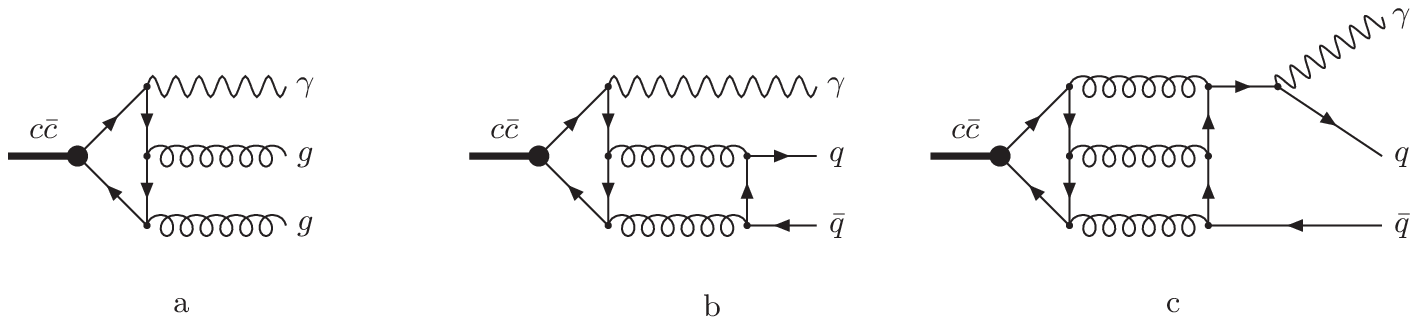}
\end{center}
  \caption{The annihilation in radiative decays of $J/\psi$: a) the gluonic channel,
  b) the gluon conversion into light quarks in the isoscalar channel, c)
  the gluon conversion into light quarks with breaking off the isospin.}
  \label{rad-gg}
\end{figure*}

Radiative decays of $J/\psi$ are considered
as a source of gluon enhanced channels relevant to searching for
glueball states predicted by QCD (see references in
\cite{Close,Anisovich}). This belief is based on the consideration
of diagrams for the annihilation of charm-quarks composing the
initial state as shown in Fig. \ref{rad-gg}a. The two-gluon state
can further covert to both gluonic hadrons and light quarks (see
Fig. \ref{rad-gg}b), that produces the mixing of pure glueballs
with the ordinary quark matter. A common belief stands that the
photon coupling to light quarks causes the suppression of
amplitude by $\alpha_s$ due to the additional exchange by the hard
gluon (see Fig. \ref{rad-gg}c), that yields the isoscalar
dominance in the radiative $J/\psi$-decays, since the contribution
due to the electromagnetic current of light quarks causing the
isospin breaking is suppressed.

However, the exclusive branching fractions, in practice, are
actually suppressed with respect to inclusive ones by an order of
magnitude, in fact. Therefore, the yields of hadrons with large
gluonic components in the radiative decays of vector charmonium
could be of the same order as the yield of ordinary hadrons due to
diagrams accompanying one in Fig. \ref{rad-gg}c. This point could
be straightforwardly checked by searching for the isospin breaking
radiative decays. Indeed,  the isospin breaking component of
diagrams with electromagnetic current coupled to light quarks
(Fig. \ref{rad-gg}c) does not interfere with the dominant
contribution of isoscalar channel (Fig. \ref{rad-gg}a and b).
Therefore, one could test the magnitude of isospin breaking
channels by investigating the transitions to isotriplet scalar and
pseudoscalar particles\footnote{Similar mechanism contributes to
hadronic decays of $J/\psi$ and $\psi^\prime$, if one considers
the virtual photons as it was studied in \cite{Zhao}, wherein a
role of isospin breaking was also emphasized.} (the isospin
$T=1$). The transition to $\pi^0$ is measured, and it is
suppressed by two orders of magnitude in comparison with decays to
isosinglet pseudoscalar particles, which can include large gluonic
components. Then, the scalar channel is of interest. It could be
represented by the $J/\psi\to a_0\gamma$ mode, for instance. In
the present paper we evaluate the branching fraction of this
channel in the framework of simple factorizing the isoscalar
heavy-light mixing and electromagnetic transition of isoscalar
vector light-quark state into the isotriplet scalar. For this
purpose in section \ref{eval} we analyze the effective couplings
of the isosinglet vector states with isotriplets ones and show
that the current data on the radiative decays allow us to extract
relevant mixing multiplied by form-factors in order to predict the
widths of $J/\psi\to a_0\gamma$. The result derived is discussed
in section \ref{conc}.

\section{Evaluation\label{eval}}

Let us introduce effective dimensionless couplings $\widetilde F$
and $F$ of vector isoscalar particle $V=\omega,\,\phi,\,J/\psi$
with the neutral isotriplet pseudoscalar $\pi^0$ and scalar $a_0$
in amplitudes of radiative decays as follows:
\begin{eqnarray}\label{amp}
    \widetilde{\mathcal{M}}(V\to\pi^0\gamma)&=&
    \frac{\sqrt{4\pi\alpha}}{2m_V}\,\widetilde{
    F}_{V\pi^0}\,\widetilde{\mathscr{F}}_{\mu\nu}\mathscr{V}^{\mu\nu},\\
    \mathcal{M}(V\to a_0\gamma)&=&
    \frac{\sqrt{4\pi\alpha}}{2m_V}\,
    F_{V a_0}\,\mathscr{F}_{\mu\nu}\mathscr{V}^{\mu\nu},
\end{eqnarray}
where $\mathscr{F}_{\mu\nu}$ is the strength tensor of
electromagnetic field $\mathscr{A}_\mu$
$$
    \mathscr{F}_{\mu\nu}=\partial_\mu\mathscr{A}_\nu-
    \partial_\nu\mathscr{A}_\mu,
$$
$\widetilde{\mathscr{F}}_{\mu\nu}$ denotes its dual tensor
$$
    \widetilde{\mathscr{F}}_{\mu\nu}=\frac{1}{2}\,
    \epsilon_{\mu\nu\alpha\beta}\mathscr{F}^{\alpha\beta},
$$
while $\mathscr{V}^{\mu\nu}$ is the strength tensor of vector
$\mathscr{V}_\mu$. For brevity we include the product of effective
electric charges $Q_V$ and $Q_{\pi^0\hskip-2pt,\hskip1pt a_0}$
into the definition of couplings $\widetilde F$ and $F$. Then, we
easily get the widths of decays
\begin{eqnarray}\label{width}
    \Gamma(V\to\pi^0\gamma)&=&
    \frac{\alpha}{3}\,\frac{k_0^3}{m_V^2}\,\widetilde{
    F}_{V\pi^0}^2,\\
    \Gamma(V\to a_0\gamma)&=&
    \frac{\alpha}{3}\,\frac{k_0^3}{m_V^2}\,F_{Va^0}^2,
\end{eqnarray}
with $k_0$ being the photon energy
$$
    k_0=\frac{m_V^2-m_{\pi^0\hskip-2pt,\hskip1pt a_0}^2}{2m_V}.
$$

The experimental values of measured branching fractions \cite{PDG}
approximately yield the effective couplings to the neutral
pion\footnote{The values in (\ref{tF})--(\ref{tF2}) reproduce
those of derived in \cite{Zhao} after appropriate rescaling in
definition: $g=\widetilde F\ \sqrt{4\pi\alpha}$.}
\begin{eqnarray}\label{tF}
    \widetilde F_{\omega\pi^0}&=&1.86\pm0.03,\\ \label{tF1}
    \widetilde F_{\phi\pi^0}&=&0.135\pm0.004,\\ \label{tF2}
    \widetilde F_{\psi\pi^0}&=&(1.81\pm0.36)\,10^{-3},
\end{eqnarray}
that spectacularly shows order by order decreasing of effective
couplings with stepping to more heavy vector state. Such the
regularity can be understood in a simple manner.

\begin{figure}[th]
\begin{center}
\includegraphics[width=5cm]{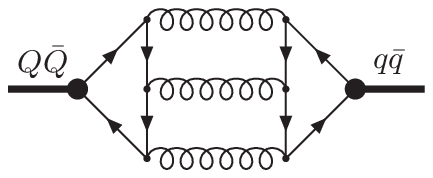}
\end{center}
  \caption{The mixing of vector isoscalar quark states in the lowest order
  of QCD coupling constant.}
  \label{rad-gg1}
\end{figure}

The diagram shown in Fig. \ref{rad-gg1} describes the mixing of
vector isoscalar quark states in the lowest order of QCD coupling
constant. Therefore, the mixing does exist, though its evaluation
is beyond the perturbation theory. The hierarchy of couplings in
(\ref{tF})--(\ref{tF2}) implies that we could represent the
transition of more heavy states $\phi$ and $J/\psi$ by the mixing
with $\omega$ and further radiative decay, i.e.
\begin{equation}\label{t-fact}
    \widetilde F_{V\pi^0}=\mathcal{K}_{V\omega}\,\widetilde
    F_{\omega\pi^0}.
\end{equation}
The $\mathcal{K}$-factors include both the mixing amplitude and
form-factors caused by the virtuality of $\omega$-propagation.
Therefore, these factors cannot be straightforwardly extracted
from the phenomenology of $\omega$--$\phi$-mixing, say.

In this way we find the $\mathcal{K}$-factors
\begin{equation}\label{tK}
    \mathcal{K}_{\phi\omega}=(7.2\pm0.2)\,10^{-2},\quad
    \mathcal{K}_{\psi\omega}=(10\pm2)\,10^{-4}.
\end{equation}

Supposing the same mechanism for the radiative decay to isotriplet
scalar meson $a_0$, we put
\begin{equation}\label{fact}
    F_{Va_0}=\mathcal{K}_{V\omega}\,
    F_{\omega a_0},
\end{equation}
while the experimental value of $\Gamma(\phi\to a_0\gamma)$ yields
\begin{equation}\label{F}
    F_{\phi a_0}=1.5\pm0.1,
\end{equation}
that results in effective value of coupling for the kinematically
forbidden decay
\begin{equation}\label{F2}
    F_{\omega a_0}=21\pm 2,
\end{equation}
as well as the quantity of interest
\begin{equation}\label{F3}
    F_{\psi a_0}=(20\pm 5)\,10^{-3},
\end{equation}
giving
\begin{equation}\label{Ga}
    \mathcal{B}r(J/\psi\to a_0\gamma)=(3.1\pm1.5)\,10^{-3},
\end{equation}
where the uncertainty is dominated by experimental one in the
measurement of radiative decays of $J/\psi$ to pion as well as
$\phi$ to $a_0$, while the theoretical uncertainty could be
additionally included \textit{ad hoc} at the level of 20\%,
characteristic for the exclusive modelling the virtual process of
mixing and form-factors.

\section{Discussion and Conclusion\label{conc}}

The only experimental data available to the moment for the
relevant process is presented by
\begin{equation}\label{expBr}
    \mathcal{B}r(J/\psi\to\gamma\pi^+\pi^-2\pi^0)=(8.3\pm
    3.1)\,10^{-3},
\end{equation}
that should be compared with expected
\begin{eqnarray}
    \mathcal{B}r(J/\psi\to\gamma a_0)\times
    \mathcal{B}r(a_0\to\eta\pi^0)\times&&\nonumber\\
    \hskip3mm\mathcal{B}r(\eta\to\pi^+\pi^-\pi^0)&=&
    (0.7\pm0.4)\,10^{-3},\nonumber
    \label{expect}
\end{eqnarray}
while the neutral modes of $\eta$ decay in such final states of
$J/\psi$ decay are not available yet. We see that the resonant
$a_0$-contribution due to the cascade decay with the
$\gamma\pi^+\pi^-2\pi^0$ final state could give about 10\% of
branching ratio summed over all modes as we expect.

The main conclusion of estimate obtained is the following: the
ordinary quark states could compose a considerable fraction of
radiative $J/\psi$-decay in contrast to usual assumption on the
enhancement of gluonic component. Moreover, the isospin breaking
effects could be essential. Thus, the simple treatment of
resonances in the scalar channel as the probable glueball states
is under question.

The critical point of estimate is the factorization of mixing
form-factors and effective couplings. This suggestion is natural.
The resulting mixing factors are small, that should support the
assumption. However, the suppression of mixing could lead to the
method suffers from small effects breaking the factorization. For
instance, the light quark states, $\omega$, $a_0$ or $\phi$ with
the hidden strangeness could get a small admixture of four-quark
states. Then, the admixture would break the factorization.
Nevertheless, we expect that the four-quark mechanism gets a
significant suppression, since in the exclusive modes one should
include the annihilation of additional quark-antiquark pair.
Therefore, we expect that the method used is able to present the
real estimate for the isospin breaking effect in the radiative
$J/\psi$-decay to $a_0$.

The author is grateful to prof. A.M.\ Zaitsev for initiating the
consideration of problem and useful discussions.

This work is partially supported by the Russian Foundation for
Basic Research, grant 07-02-00417-a.

\end{document}